\documentclass{iau}
\usepackage{graphicx} 

\title[IAUS291.~~Pulsar force-free magnetosphere revisited] %% short title %%
{Global structure of the pulsar force-free magnetosphere revisited} %% full title %%

\author[S.~Petrova]  %% short author list %%
{Svetlana Petrova}

\affiliation{Institute of Radio Astronomy, \\Chervonopraporna Str., 4, 61002 Kharkov, Ukraine \\ email: {\tt petrova@ri.kharkov.ua}}

\pubyear{2012}
\volume{291}  %% insert here IAU Symposium No.
\jname{\mbox{Neutron Stars and Pulsars: Challenges and Opportunities after 80 years}}
\editors{J.~van Leeuwen, ed.} 
\begin{document}

\maketitle

%% -- Abstract ----------------------------------
\begin{abstract}
A new model of the pulsar force-free magnetosphere is suggested, which includes the presence of the polar, outer and slot gaps. It is based on a new exact solution of the pulsar equation in the form of an offset monopole and the resultant split-offset monopole scheme.
\keywords{MHD, plasmas, pulsars: general, stars: neutron, stars: magnetic fields}
\end{abstract}

% add below any authors, subjects and objects for indexing 
%   add more lines if necessary
%   but leave all lines commented out
%\index[author]{LastName1, Initials|textbf}
%\index[author]{LastName2, Initials|textbf}
%\index[subject]{Keyword1}
%\index[subject]{Keyword2}
%\index[object]{Object1}
%\index[object]{Object2}

\firstsection % if your document starts with a section,
\section{Introduction}

The primordial dipolar structure of the pulsar magnetic field is modified by the plasma originating in the pulsar magnetosphere. The problem of a self-consistent description of fields and currents in the pulsar magnetosphere was formulated in the form of a well-known pulsar equation (\cite[Michel 1973]{m73}). The basic model is that of a stationary axisymmetric force-free dipole, where the magnetic and rotational axes are aligned, the electromagnetic forces balanced and the particle inertia is ignored.

The problem lies in guessing a current distribution which makes the self-consistent magnetic field obey the physically meaningful boundary conditions. Such a current function along with the consequent magnetic field structure was first simulated numerically in the pioneering work of \cite[Contopoulos et al. (1999)]{ckf99}. Later on these results were developed in a number of aspects (for a review see \cite[Spitkovsky]{s12}, this volume). All the subsequent studies confirmed the formal validity of the original simulations of \cite[Contopoulos et al. (1999)]{ckf99}, however the physical meaning of the current function obtained is still questionable, since it implies that a part of the return current should flow through the polar gap.

We suggest to revise the commonly used set of boundary conditions in the pulsar force-free problem so as to allow for the presence of the plasma-producing gaps. Only in this case the treatment of the pulsar magnetosphere can be truly self-consistent.

\section{Split-offset monopole scheme of a force-free dipole}

{\underline{\it New exact solution of the pulsar equation}}. An exact analytic solution of the pulsar equation is known only for a monopole located at the centre of a neutron star (\cite[Michel 1973]{m73}). We have found (\cite[Petrova 2012a]{p12a}) that the magnetic flux function of a monopole offset by a distance $a$ along the $z$-axis of the cylindrical coordinate system $(\rho,\phi,z)$,
\begin{equation}
f=f_0\left[1-\frac{z-a}{\sqrt{(z-a)^2+\rho^2}}\right],
\label{eq1}
\end{equation}
also satisfies the pulsar equation, in which case the current function $A$ and the velocity of differential rotation $\Omega$ are related as
\begin{equation}
A=\Omega f\left(2-f/f_0\right).
\label{eq2}
\end{equation}
As $r\equiv\sqrt{\rho^2+z^2}\to\infty$, Eq.~(\ref{eq1}) approaches the flux function of a centered monopole. At finite large distances, $r\gg 1,$ the flux function~(\ref{eq1}) is an infinite series over the centered multipoles and presents a generalized multipolar solution of the pulsar equation.

{\underline{\it Implications of the offset monopole solution}}. At infinity, the field of a force-free dipole is generally believed to be well represented by the classical split monopole scheme. At finite distances, however, the scheme does not allow for the consequences of essentially dipolar features. The analogy with an offset monopole enables to account for the outer gap in the pulsar magnetosphere. As the outer gap forms at the intersection of the null line with open magnetic field lines and is expected to lie entirely within the light cylinder, beyond the light cylinder the null line should coincide with a certain magnetic field line. Given that the outer gap is the place of passage of the return current, it is the critical field line that coincides with the null line, in which case both go parallel to the equator at a certain altitude above it, similarly to the case of an offset monopole.

\begin{figure}[t]
 \begin{center}
  \includegraphics[width=3.4in]{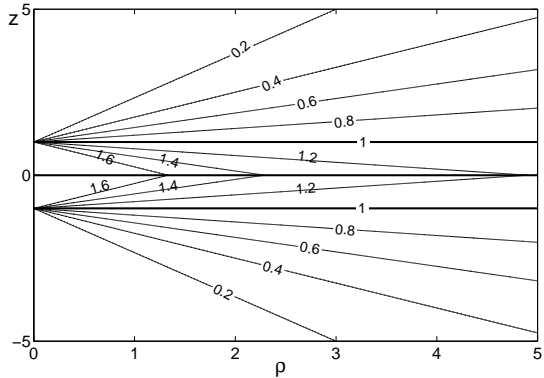}
  \caption{Split-offset monopole scheme. The level lines of the magnetic flux function are presented together with the level values. Thick lines show possible current sheet locations.} 
    \label{fig1}
 \end{center}
 \end{figure}

{\underline{\it Basics of the split-offset monopole scheme}}. Based on the above considerations, we suggest that beyond the light cylinder the pulsar force-free field is better represented by the split-offset monopole scheme composed of the two symmetrically offset monopoles of opposite polarity (see Fig.\,\ref{fig1}). With Eqs.~(\ref{eq1})-(\ref{eq2}) it is easy to show that the condition $B^2-E^2$ (where $B$ and $E$ are the magnetic and electric field strengths, respectively) is fulfilled everywhere and for any pair $(A,\Omega)$ satisfying Eq.~(\ref{eq2}). Thus, the force-free approximation is valid. Furthermore, for the three horizontal lines in Fig.\,\ref{fig1} (i.e. the two null (critical) lines and the equator) the equilibrium condition $\mathrm{d}(B^2-E^2)/\mathrm{d}z=0$ (e.g., \cite[Lyubarsky 1990]{l90}) is valid for any relevant pair $(A,\Omega)$. Correspondingly, the four regions in Fig.\,\ref{fig1} bounded by the three above mentioned horizontal lines may have different $(A,\Omega)$.

{\underline{\it Current circuit configurations}}. In the split-offset monopole scheme, the current circuit may contain from one to three current sheets located along the null (critical) lines and the equator. In the simplest case we have $A=\Omega=0$ in the equatorial region between the lines $f=f_0$ and $A\neq 0$ between the magnetic axis and these lines. Then the symmetric current sheets along the lines $f=f_0$ close the current circuit in each hemisphere. If in the equatorial region $A\neq 0$ as well, the regions on both sides of the lines $f=f_0$ may join without current sheets, and the return current flows in the equatorial current sheet and along the equatorial lines entering the sheet. Note that a similar structure of the equatorial region beyond the light cylinder is characteristic of the dipolar force-free magnetosphere simulated by \cite[Gruzinov (2011a,b)]{g11a,g11b}.

\begin{figure}[t]
 \begin{center}
 
 \vspace{23mm}
 
  \includegraphics[width=2.2in]{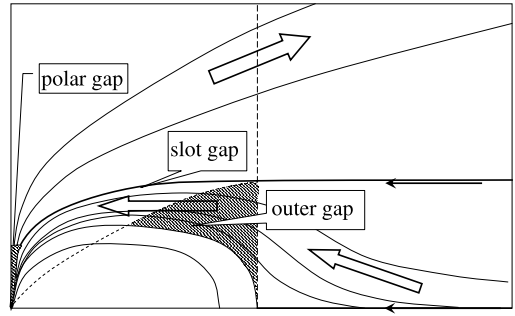}
  
  \vspace{-10mm}
  
    \caption{Schematic structure of the pulsar force-free magnetosphere allowing for the plasma-producing gaps. The dashed line represents the light cylinder boundary; the dotted line delineates the null line; the thick line corresponds to the critical line; arrows show the current flow directions.} 
    \label{fig2}
 \end{center}
 \end{figure}

\section{Implications of the split-offset monopole model}

Based on the split-offset monopole model, the global structure of the pulsar force-free magnetosphere seems to look as follows (see Fig.\,\ref{fig2}). The polar and outer gaps control different bundles of open field lines divided by the critical line, and the two gaps are adjusted by the slot gap located between them. Note that a similar configuration of the coexisting gaps was recently obtained in \cite[Yuki \& Shibata (2012)]{ys12} by means of numerical simulations of the plasma particle motions in the pulsar magnetosphere. The direct current flows through the polar and slot gaps and returns to the neutron star through the outer gap. In the outer and slot gaps, the longitudinal current may change along a field line because of the trans-field currents near the light cylinder (\cite[Petrova 2012a]{p12a}).

Our schematic model is believed to be a proper basis for detailed analytic and numerical studies of the pulsar force-free magnetosphere. We have already started a systematic analytic description of the model (\cite[Petrova 2012b]{p12b}). The first results concern the region near the magnetic axis and testify to the distinction of the force-free field at the top of the polar gap from that of a pure dipole. This is attributed to the action of the transverse current flowing at the neutron star surface and closing the pulsar current circuit.

\section*{Acknowledgements}

I acknowledge partial support by the IAU Travel Grant. The work is partially supported by the grant of the President of Ukraine (the project of the State Fund for Fundamental Research No. F35/554-2011).

\end{document}